\documentclass[conference]{IEEEtran}
\IEEEoverridecommandlockouts
\usepackage{cite}
\usepackage{amsmath,amssymb,amsfonts}
\usepackage{algorithmic}
\usepackage{graphicx}
\usepackage{textcomp}
\usepackage{xcolor}
\usepackage{booktabs}
\usepackage{tabularx}
\usepackage[table]{xcolor}
\definecolor{headerblue}{RGB}{220,230,245}
\definecolor{sectiongray}{RGB}{240,240,240}
\definecolor{bestgreen}{RGB}{0,110,0}
\definecolor{bestblue}{RGB}{0,70,170}
\usepackage{newunicodechar}
\usepackage[most]{tcolorbox}
\usepackage{subcaption}
\usepackage{multirow}
\def\BibTeX{{\rm B\kern-.05em{\sc i\kern-.025em b}\kern-.08em
    T\kern-.1667em\lower.7ex\hbox{E}\kern-.125emX}}
\begin{document}

\title{Candor-LR: A Dyadic Conversational Dataset for Audio-Visual Speech Recognition\\
\thanks{This publication emanates from research supported by Taighde Éireann – Research Ireland, Grant number 22/FFP-A/11059.}
}

\author{
\IEEEauthorblockN{
Rishabh Jain \quad
Aristeidis Papadopoulos \quad
Zhaofeng Lin \quad
Naomi Harte
}
\IEEEauthorblockA{
\textit{Sigmedia Group, School of Engineering, Trinity College Dublin, Ireland} \\
\{rijain, papadoar, linzh, nharte\}@tcd.ie
}
}





\maketitle

\begin{abstract}

Current audio-visual speech recognition (AVSR) benchmarks, like LRS3, rely heavily on clean, scripted and rehearsed speech. They fail to reflect the complexity of natural conversation, which involves overlapping speech, spontaneous turn-taking, unscripted vocabulary and variable acoustic conditions. To shift the field toward realistic dialogue, we introduce Candor-LR, a conversational benchmark derived from the CANDOR corpus of 1,656 natural dyadic videoconferences. Our custom data preparation pipeline yields 713.5, 10.1, and 60.1 hours of training, validation, and test data, respectively. Evaluating pretrained AVSR models on Candor-LR reveals that audio-only accuracy drops sharply compared to LRS3, but visual cues compensate effectively, driving much larger performance gains on Candor-LR than on LRS3. Furthermore, training on this corpus significantly improves cross-domain robustness under both clean and noisy conditions, as its realistic conversational data captures broader audio-video features. We open-source our pipeline to ensure reproducibility, establishing Candor-LR as a challenging benchmark for conversational AVSR.
\end{abstract}

\begin{IEEEkeywords}
Audio-visual speech recognition, CANDOR dataset, benchmark, noise robustness, video conferencing data
\end{IEEEkeywords}

\section{Introduction}

Audio-visual speech recognition has made remarkable progress on benchmark datasets \cite{prajwal24_interspeech, ahn24_interspeech, avformer, kit2025phoneme, mms, rouditchenko24_interspeech, MARC}, with AV-HuBERT \cite{avhubert} reaching 1.47\% WER on LRS3 \cite{lrs3} and Auto-AVSR \cite{autoavsr} reaching 0.90\%. However, these gains are largely measured on scripted datasets such as LRS2 \cite{lrs2} and LRS3 \cite{lrs3}. These datasets feature professional speakers delivering well-articulated speech under relatively consistent recording conditions and with high-quality microphones, which differ substantially from everyday conversational settings. They contain no overlapping speech, no spontaneous interruptions, and little acoustic variability. As a result, strong benchmark performance does not necessarily guarantee robustness in real conversational settings.

Real conversation is fundamentally different. People interrupt each other, turn away mid-sentence, speak over one another, and deal with variable room acoustics \cite{WAGNER20151, LINKE2025101738}. These conversational behaviors are largely absent from current AVSR benchmarks, which means strong performance on scripted speech does not necessarily transfer to everyday dialogue. Recent work has also questioned what current systems actually learn from visual speech. State-of-the-art VSR and AVSR models often rely heavily on dataset patterns and language modeling rather than genuinely stronger visual representations \cite{lip_reading_gap,jain2026hypeinsightrethinkinglarge}. Other analyses further show that low WER does not necessarily imply strong use of visual information, suggesting that current benchmarks can overestimate true visual speech understanding \cite{lin2025uncovering}.

Recent AVSR datasets have started to shift towards more realistic conditions \cite{multivsr, ci-asvr, friend_mmc, chinese_lips}. WildVSR shows that models tuned to LRS3 degrade sharply on a harder in-the-wild test set \cite{djilali2024vsr}. LRS-VoxMM \cite{lrs_voxmm} further shows that visual cues become more important as audio quality degrades, while the addition of noise and reverberation creates substantially more challenging AVSR evaluation conditions. Cocktail-party AVSR \cite{nguyen25b_interspeech, cocktail_benchmark} adds overlapping speakers, interfering talkers, and silent-face segments, showing that current models can deteriorate sharply in realistic multi-speaker settings. In-the-wild conversational benchmarks \cite{Ma2022, djilali2024vsr} further show that AVSR becomes more difficult once audio quality degrades, and recent work on video conferencing \cite{av_conference, MISP} shows that platform-induced distortions and speech enhancement artifacts can also cause severe performance collapse. Together, these datasets are valuable, but they still do not provide the scale and realism needed to move AVSR beyond broadcast-style evaluation and toward a more realistic conversational style paradigm \cite{Nguyen_2026_CVPR}. There remains a clear need for a benchmark that can support both training and testing in realistic dyadic settings and to drive progress towards conversational AVSR.

To address this gap, we present Candor-LR, a benchmark derived from the CANDOR \cite{candor} corpus of 1,656 natural dyadic videoconference conversations. We develop a custom pipeline to convert CANDOR into AVSR-ready data using Speechmatics \cite{speechmatics} word-level time-aligned transcripts. Candor-LR contains 713.5 hours of AVSR-formatted data for training, 10.1 hours for validation, and 60.1 hours for testing. Unlike scripted benchmarks, Candor-LR is built from spontaneous dyadic conversation, including turn-taking, unscripted vocabulary, and realistic acoustics, with final segments extracted as single-speaker phrases from per-speaker tracks. Our experiments show that AVSR systems trained on scripted data perform poorly on real conversation. Visual cues matter more in these realistic settings, and training on Candor-LR improves robustness on both clean and noisy speech. Our results highlight a key limitation in AVSR: models trained on scripted datasets do not generalize well to real dialogue. Rather than offering only an incremental step in AVSR research, our work aims to reposition AVSR around natural conversational settings that better reflect real-world use. We also open-source the pipeline on GitHub for converting the original CANDOR corpus\footnote{Users must obtain their own license from the CANDOR authors \cite{candor}; we do not distribute the dataset and are not responsible for licensing.} into AVSR-ready format.

\section{CANDOR Cleaning for AVSR}
The CANDOR dataset \cite{candor} has 1,656 natural dyadic conversations. It contains over 850 hours of video and 7 million words. Most AVSR datasets, like LRS2 or LRS3, only show one person talking for a few seconds. In contrast, CANDOR contains two-speaker sessions of approximately 26-30 minutes. It was originally developed for research on natural conversation and social interaction, not audio-visual speech recognition. As the conversations are unscripted, they include overlapping speech, a wide range of vocabulary, and frequent head movements. Off-the-shelf AVSR preprocessing tools are designed for broadcast, single-speaker recordings and are not suited for conversational data of this kind. We therefore developed a custom pipeline to convert CANDOR into a format suitable for AVSR training and evaluation. The original directory structure of the CANDOR dataset is shown in Figure \ref{fig:candor-structure}.

\begin{figure}[t]
\centering
\begin{tcolorbox}[
  colback=white,
  colframe=black!60,
  boxrule=0.5pt,
  arc=2pt,
  left=6pt,
  right=6pt,
  top=4pt,
  bottom=4pt,
  width=\linewidth
]
\renewcommand{\arraystretch}{1.0}
\footnotesize 
\begin{tabular}{@{} l l @{}}
\textbf{\{session\_id\}/} & \\
$\vdash$ \textbf{processed/} & \\
\textbar~~$\vdash$ \texttt{\{session\_id\}.mp4} & \color{gray}{\# Combined video (640x240)} \\
\textbar~~$\vdash$ \texttt{\{session\_id\}.mp3} & \color{gray}{\# Combined stereo audio} \\
\textbar~~$\vdash$ \texttt{\{user\_id\_1\}.mp4}  & \color{gray}{\# Spkr 1 video (320x240, 30fps)} \\
\textbar~~$\vdash$ \texttt{\{user\_id\_2\}.mp4}  & \color{gray}{\# Spkr 2 video (320x240, 30fps)} \\
\textbar~~$\vdash$ \texttt{channel\_map.json}   & \color{gray}{\# Maps L/R audio to user IDs} \\
\textbar~~$\llcorner$ \texttt{metadata.json}      & \color{gray}{\# Session info and sync details} \\
$\vdash$ \textbf{transcription/} & \\
\textbar~~$\vdash$ \texttt{transcribe\_out.json} & \color{gray}{\# Raw AWS ASR output} \\
\textbar~~$\vdash$ \texttt{transcript\_aud.csv}  & \color{gray}{\# Turn-level transcripts} \\
\textbar~~$\vdash$ \texttt{transcript\_back.csv} & \color{gray}{\# Backchannel-aware model} \\
\textbar~~$\llcorner$ \texttt{transcript\_cliff.csv} & \color{gray}{\# Sentence-boundary model} \\
$\llcorner$ \textbf{raw/} & \color{gray}{\# Original recordings} \\
\end{tabular}
\end{tcolorbox}
\vspace{-0.8em}
\caption{Directory structure of CANDOR dataset.}
\label{fig:candor-structure}
\vspace{-1.5em}
\end{figure}

\subsection{Transcript Source: Speechmatics over Originals}

The CANDOR corpus includes automatic transcripts created with the AWS Transcribe API. Based on these, Reece et al. \cite{candor} provided three ways to split the dialogue into speaker turns: Audiophile (simple speaker switches), Cliffhanger (sentence boundaries), and Backbiter (which accounts for listener feedback). While these are useful for understanding the conversation, they do not include word-level timestamps. For AVSR training, we need millisecond-level precision to match the audio to the video, which these original transcripts do not provide.
To address this limitation, we use word-level transcripts generated by the Speechmatics ASR system, as produced by Russell et al. \cite{RUSSELL2024100163}. Speechmatics was selected for its strong performance on conversational speech and its ability to accurately capture non-lexical tokens (like "uh" or "um"). Most importantly, it provides per-word confidence scores, punctuation, and millisecond-precision start and end timestamps for every word. These timestamps are used to drive all timing and segmentation in our pipeline. The organization of these Speechmatics outputs is shown in Figure \ref{fig:speechmatics-structure}.

\begin{figure}[t]
\centering

\begin{tcolorbox}[
  colback=white,
  colframe=black!60,
  boxrule=0.5pt,
  arc=2pt,
  left=2pt,
  right=2pt,
  top=2pt,
  bottom=2pt
]

{\footnotesize

\textbf{candor\_speechmatics/}\\[-2pt]

\hspace*{0.6em}\texttt{\{session\_id\}\_0.json} \hspace{1em} \# speaker 0 word-level JSON\\
\hspace*{0.6em}\texttt{\{session\_id\}\_1.json} \hspace{1em} \# speaker 1 word-level JSON\\
\hspace*{0.6em}\texttt{\{session\_id\}.TextGrid} \hspace{1em} \# combined Praat alignment

}
\end{tcolorbox}
\vspace{-0.8em}
\caption{Directory structure of Speechmatics-ASR outputs.}
\label{fig:speechmatics-structure}
\vspace{-1.5em}
\end{figure}

\subsection{Speaker-Video Mapping}

CANDOR's dyadic nature introduces an alignment challenge: speaker videos are identified by unique User IDs (e.g., \texttt{\{user\_id\}.mp4}), whereas the Speechmatics transcripts are indexed by audio channel number (e.g., \texttt{\_0.json} and \texttt{\_1.json}). Without a direct link, the transcript for a specific speaker cannot be automatically identified. To resolve this, we utilize the \texttt{channel\_map.json} file (see Figure \ref{fig:candor-structure}). This metadata file maps the Left (L) and Right (R) audio channels to their corresponding speaker User IDs. Specifically, we map channel \texttt{0} to the \texttt{L} video and channel \texttt{1} to the \texttt{R} video. This ensures that each word-level transcript is correctly synchronized with the appropriate speaker's video stream.

\subsection{Phrase-Level Temporal Segmentation}
CANDOR sessions average 26 minutes, necessitating segmentation into 2–5 second clips (similar to LRS3) suitable for AVSR training. We group words from the Speechmatics output into phrases, breaking at punctuation or inter-word gaps over 0.5 seconds. This method avoids the context loss of word-level segments (0.2-1.0s) and the silence of turn-level segments (0.5-84s), aligning our data with LRS2/LRS3 standards. On the full corpus, this process yields clips ranging from 0.8 to 5.0 seconds, with a mean duration of 3.6 seconds. 

\subsection{Face Detection and Mouth ROI Extraction}
We applied RetinaFace \cite{RF} frame-by-frame to each video clip with detection threshold of 0.8. Following successful face detection, 68 facial landmarks were extracted, and the mouth region (landmarks 48-68) was cropped and resized to 96×96 pixels at 25 fps (downsampled from the original 30 fps for consistency with LRS2/LRS3 preprocessing). To reduce inter-frame jitter caused by spontaneous head movement, a 3-frame moving average is applied to the landmark positions before cropping.

\subsection{Audio Processing and Synchronization}
Audio segments are extracted directly from the individual per-speaker audio tracks using the same phrase-level timestamps derived from the Speechmatics word alignments. Extracting from dedicated per-speaker tracks, rather than the combined stereo recording, eliminates cross-talk from the other conversation participant. Each segment is converted from stereo to mono, resampled to 16 kHz, and saved as a lossless WAV file. Because audio and video are extracted using the same timestamps, frame-perfect synchronization is guaranteed without requiring any additional alignment step.

\subsection{Text Normalization and Quality Filtering}
To ensure compatibility with standard AVSR frameworks like AV-HuBERT \cite{avhubert} and Auto-AVSR \cite{autoavsr}, transcripts are normalized by removing punctuation, collapsing redundant whitespace, and converting text to lowercase. Remaining disfluency markers from the Speechmatics output are removed to maintain consistent training labels. We further apply three sequential filters at the phrase level to remove artifacts unsuitable for AVSR. First, a duration filter discards clips shorter than 800 ms, as these lack sufficient visual context. Second, a word count filter removes utterances with fewer than two words. Finally, a filler word filter removes phrases consisting entirely of non-lexical tokens (e.g., "uh," "um," "mhm"); phrases containing at least one lexical word are retained. These filters are applied to the full dataset prior to any splitting, ensuring consistent quality across training, validation, and test sets, and any clip failing them is discarded alongside its corresponding audio and transcript.

\subsection{Data Splitting and Output Format}
The dataset is split into training, validation, and testing sets using a speaker-aware partitioning approach. To construct a truly unseen test set, we identify speakers who appear in exactly one session across the entire corpus: a total of 701 such speakers accounting for approximately 164.9 hours of audio. These single-occurrence speakers are reserved exclusively for the test split, ensuring that no speaker in the test set is encountered during training or validation. The remaining multi-session speakers are partitioned into training and validation sets using a fixed random seed, with conversational contexts never overlapping across splits. These splits are saved as canonical ID files (\texttt{candor-train.id}, \texttt{candor-valid.id}, \texttt{candor-test.id}) to ensure reproducibility. The complete pipeline explanation, hyperparameters used and source code are available on our GitHub.\footnote{https://github.com/rishabhjain16/lipreading-data-guide/tree/main/Candor} The prepared AVSR dataset is referred to as \textbf{Candor-LR} throughout the rest of this paper.

\section{Candor-LR}
Candor-LR contains \textbf{783.7} hours (787,670 utterances) of AVSR-formatted data. This comprises approximately \textbf{713.5 hours} (718,648 utterances) for training, \textbf{10.1} hours (10,304 utterances) for validation, and \textbf{60.1} hours (58,718 utterances) for testing, corresponding to 91.2\%, 1.3\%, and 7.5\% of the total data respectively. Across all splits, Candor-LR contains 1,554 speakers with valid demographic metadata (1,350 train, 44 validation, 160 test), of which 1,521 are unique after accounting for 33 speakers overlapping between training and validation. Speakers in the test set are disjoint from those in the training and validation sets. For comparison, LRS3 consists of 9,506 videos (5,090 pre-train, 4,004 trainval, 412 test) totaling 151,819 utterances (118,516 pre-train, 31,982 trainval, 1,321 test) \cite{lrs3}. The LRS3 test set contains only 1,321 utterances ($\approx1$ hour), compared to Candor-LR's substantially larger test set of 58,718 utterances (60.1 hours).

\begin{figure}[t]
    \centering
    \includegraphics[width=1\columnwidth]{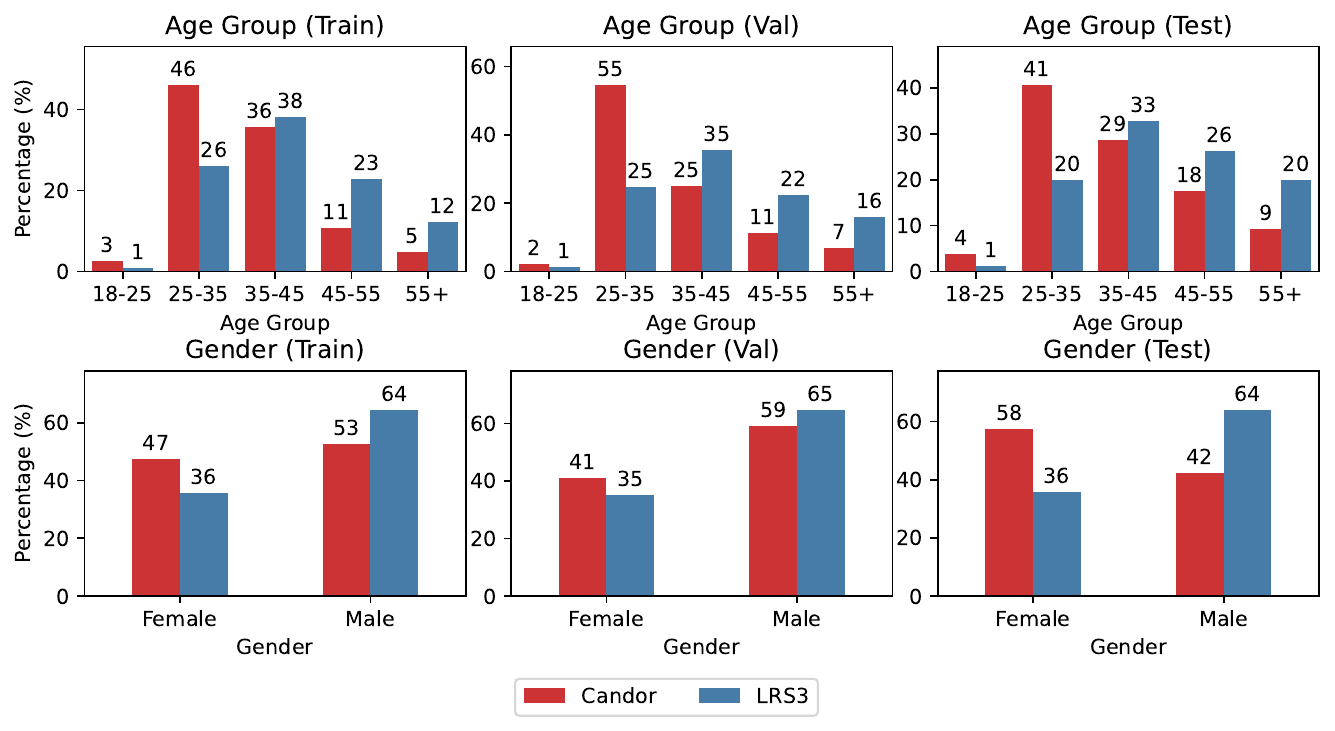}
    \caption{Age and gender distributions for Candor-LR and LRS3 estimated using UniFace. Candor-LR shows improved demographic balance across both dimensions.}
    \label{fig:stats}
    \vspace{-1.5em}
\end{figure}

The authors of CANDOR release no speaker-level demographic metadata alongside the corpus. Therefore to assess the demographic diversity of Candor-LR, we compare its speaker-level age and gender distribution against LRS3. We extract metadata using the open-source UniFace toolkit\footnote{https://github.com/yakhyo/uniface}, which estimates age and gender from video frames following the same methodology as Lin et al.~\cite{multivsr_lrs3}. Figure~\ref{fig:stats} shows the percentage-wise distributions. 

\textbf{Demographics:} Using UniFace metadata for age and gender estimation, Candor-LR shows a larger proportion of speakers in the 25-35 age bracket, 46\% of training speakers are aged 25–35, compared to 26\% in LRS3. LRS3 has higher concentration in older groups: 23\% aged 45–55 and 12\% aged 55+, compared to Candor-LR's 11\% and 5\%, respectively. For gender, Candor-LR is more balanced in training (47\% female vs. 36\% in LRS3), narrowing the gap by nearly 10 percentage points. This gender balance is even more pronounced in the test split, where Candor-LR has a slight female majority (57\% female) while LRS3 remains male-skewed (64\% male). Overall, Candor-LR offers balance across both age and gender.

\section{Experiment and Results}

\subsection{Benchmarking Candor-LR with pretrained AVSR Models}
\label{sec:benchmarking}

\begin{table*}[t]
\centering
\caption{WER (\%) performance comparison across modalities (VO, AO, and AV) on LRS3 and Candor-LR datasets.
}
\label{tab:wer-detailed}
\small
\renewcommand{\arraystretch}{1.2}

\begin{tabularx}{\textwidth}{|l|c|c| *{3}{>{\centering\arraybackslash}X|}c| *{3}{>{\centering\arraybackslash}X|}c|}
\hline
 \rowcolor{headerblue}& & & \multicolumn{4}{c|}{\textbf{LRS3}} & \multicolumn{4}{c|}{\textbf{Candor-LR}} \\ \cline{4-11}
\rowcolor{headerblue}\textbf{Model} & \textbf{Learning Type} & \textbf{Train Hours} & \textbf{VO} & \textbf{AO} & \textbf{AV} & \textbf{$\Delta$} & \textbf{VO} & \textbf{AO} & \textbf{AV} & \textbf{$\Delta$} \\ \hline
AV-HuBERT      & Self-Supervised  & 1,759 & 28.69 & 1.95 & 1.47 & +0.48 &  78.35 & 24.32 & 22.76 & +1.56 \\ \hline
Llama-AVSR     & LLM-based Decoder & 1,759 & 26.20 & 0.74 & 0.79 & -0.05 & 111.26 & 16.92 & 16.92 &  0.00 \\ \hline
Auto-AVSR-S    & Supervised       & 1,759 & 24.60 & 1.00 & --   & --    &  75.95 & 16.83 & --   & --    \\ \hline
Auto-AVSR-L    & Supervised       & 3,448 & 19.10 & 1.00 & 0.90 & +0.10 &  70.48 & 17.26 & 16.69 & +0.57 \\ \hline
\end{tabularx}
\begin{flushleft}
\scriptsize \textbf{Note:} $\Delta = AO - AV$, a positive delta indicates that the inclusion of visual information reduced the WER. All models were trained on LRS3 and VoxCeleb2, except Auto-AVSR-L, which included LRW, LRS2, LRS3, AVSpeech, and VoxCeleb2. Auto-AVSR-S AV omitted due to lack of public checkpoint.
\end{flushleft}
\vspace{-1.5em}
\end{table*}

To assess zero-shot cross-domain generalization, we evaluate three pretrained AVSR models on Candor-LR without downstream finetuning. \textbf{AV-HuBERT} \cite{avhubert} is a state-of-the-art self-supervised representation learning framework that learns by capturing the correlation between audio and visual streams. The model encodes masked audio and image sequences via a hybrid ResNet-Transformer architecture to predict multimodal hidden units. For our work, we use the checkpoint trained on 1,759-hour LRS3 and VoxCeleb2 \cite{vc2} datasets. \textbf{Llama-AVSR} \cite{llama_avsr} is a multimodal framework that utilizes a frozen large language model (LLM) \cite{llama2} as an auto-regressive text decoder. It extracts features using a frozen pretrained encoder \cite{avhubert,whisper} and maps them into the LLM's text space using modality-specific low-rank adaptation (LoRA) \cite{hu2022lora} projectors. It is trained on the same 1,759-hour corpus. Finally, \textbf{Auto-AVSR} \cite{autoavsr} is a supervised approach designed to scale AVSR by leveraging automatically generated labels. To bypass the expensive process of manual annotation, this framework utilizes powerful pretrained audio-only ASR models to transcribe massive unlabeled datasets. We use Auto-AVSR in two configurations: \textbf{Auto-AVSR-S}, trained on the 1,759-hour corpus, and \textbf{Auto-AVSR-L}, trained on a 3,448-hour dataset across five corpora (LRW \cite{lrw}, LRS2, LRS3, AVSpeech \cite{AVSpeech}, and VoxCeleb2 \cite{vc2}).

Table~\ref{tab:wer-detailed} shows WER performance across video-only (VO), audio-only (AO), and audio-visual (AV) modalities. On LRS3, all models achieve strong AO performance (WER: 0.74--1.95\%), reflecting the dataset's scripted and acoustically controlled nature. On Candor-LR, performance drops significantly (AO WER: 16.83–24.32\%), showing that dyadic conversational speech in Candor-LR provides a much more difficult benchmark. Llama-AVSR maintains the best AO performance on both datasets (LRS3: 0.74, Candor-LR: 16.92).

VO performance is consistently poor on Candor-LR. On LRS3, Auto-AVSR-L achieves the best VO performance (19.10\%) compared to AV-HuBERT (28.69\%) and Llama-AVSR (26.20\%), suggesting that larger and more diverse training data improves VO recognition on scripted speech. However, this advantage narrows on Candor-LR (Auto-AVSR-L: 70.48\% vs AV-HuBERT: 78.35\%), indicating that VO recognition is especially brittle for natural conversational speech, where spontaneous articulation creates a much more difficult setting than scripted speech.

The AV modality provides only marginal gains over AO on both LRS3 and Candor-LR. On LRS3, this is unsurprising given AO performance is already near-perfect, leaving little room for visual cues to contribute. On Candor-LR, absolute AV gains remain small ($\Delta \le +1.56$), and given the already high AO WER, it is difficult to draw conclusions about the contribution of visual information from these zero-shot results alone. Llama-AVSR shows no visual benefit on either dataset ($\Delta = 0.00$ on Candor-LR, $-0.05$ on LRS3), suggesting its LLM decoder does not effectively leverage visual features. Auto-AVSR-L achieves the best overall Candor-LR performance (AV WER: 16.69\%, $\Delta = +0.57$), yet its AO WER on Candor-LR (17.26\%) is slightly higher than Auto-AVSR-S (16.83\%), despite being trained on nearly double the data. This suggests that additional training data from diverse scripted sources does not necessarily improve AO performance on naturalistic conversational speech.

\subsection{Cross-Domain Generalization and Multi-Dataset Training}

\begin{table}[t]
\centering
\caption{VO, AO, and AV WER (\%) performance comparison across training sets (AV-HuBERT).}
\label{tab:vo_results}
\footnotesize
\setlength{\tabcolsep}{6pt}
\renewcommand{\arraystretch}{1.2}

\begin{tabularx}{\columnwidth}{|l| *{4}{>{\centering\arraybackslash}X|}}
\hline
\rowcolor{headerblue}
\textbf{Test Set} &
\textbf{VO} &
\textbf{AO} &
\textbf{AV} &
\textbf{$\Delta$} \\
\hline

\rowcolor{sectiongray}
\multicolumn{5}{|c|}{\textbf{Trained on LRS3: 433 Hours}} \\
\hline
LRS3      & 28.60 & 2.99 & 1.65 & \textbf{\textcolor{bestblue}{+1.34}} \\
LRS2      & 38.00 & 9.43 & 7.25 & \textbf{\textcolor{bestblue}{+2.18}} \\
Candor-LR & 78.35 & 26.63 & 24.86 & +1.77 \\
\hline

\rowcolor{sectiongray}
\multicolumn{5}{|c|}{\textbf{Trained on LRS2+LRS3: 657 Hours}} \\
\hline
LRS3      & 28.29 & 2.60 & 1.60 & +1.00 \\
LRS2      & 26.58 & \textbf{\textcolor{bestgreen}{3.63}} & 2.85 & +0.78 \\
Candor-LR & 76.83 & 22.72 & 21.32 & +1.40 \\
\hline

\rowcolor{sectiongray}
\multicolumn{5}{|c|}{\textbf{Trained on Candor-LR: 713 Hours}} \\
\hline
LRS3      & 39.03 & 6.88 & 6.02 & +0.86 \\
LRS2      & 47.09 & 10.72 & 9.15 & +1.57 \\
Candor-LR & 69.60 & \textbf{\textcolor{bestgreen}{14.81}} & 10.50 & +4.31 \\
\hline

\rowcolor{sectiongray}
\multicolumn{5}{|c|}{\textbf{Trained on LRS2+LRS3+Candor-LR: 1370 Hours}} \\
\hline
LRS3 &
\textbf{\textcolor{bestgreen}{27.89}} &
\textbf{\textcolor{bestgreen}{2.42}} &
\textbf{\textcolor{bestgreen}{1.42}} &
+1.00 \\
LRS2 &
\textbf{\textcolor{bestgreen}{24.47}} &
3.76 &
\textbf{\textcolor{bestgreen}{2.54}} &
+1.22 \\
Candor-LR &
\textbf{\textcolor{bestgreen}{69.50}} &
16.37 &
\textbf{\textcolor{bestgreen}{9.83}} &
\textbf{\textcolor{bestblue}{+6.54}} \\
\hline

\end{tabularx}

\begin{flushleft}
\scriptsize \textbf{Note:} $\Delta = AO - AV$; a positive $\Delta$ indicates that the inclusion of the visual modality reduced the WER. All models in this table were \textbf{trained without additional noise augmentation}, in contrast to the best AV-HuBERT checkpoint reported in Table~\ref{tab:wer-detailed}.
\end{flushleft}
\vspace{-1.5em}
\end{table}

We further analyze AV-HuBERT under different training data configurations to examine how Candor-LR affects cross-training generalization. AV-HuBERT was chosen for this because it has a comparatively lightweight decoding setup and lower computational requirements for training than both Auto-AVSR and Llama-AVSR. It also avoids the additional complexity introduced by an LLM-based decoder, while also showing the largest visual benefit on Candor-LR in Section~\ref{sec:benchmarking}. Table~\ref{tab:vo_results} reports VO, AO, and AV WER when the model is trained on LRS3 only, LRS2+LRS3, Candor-LR only, and the combined LRS2+LRS3+Candor-LR setup.

VO performance is consistently poor on Candor-LR across all training setups, showing that visual information alone is insufficient for this dataset. AO performance varies significantly by training data, ranging from 26.63\% (LRS3-only) down to 14.81\% (Candor-LR-only), with the combined setup falling in between at 16.37\%. Training only on LRS3 gives strong in-domain performance on LRS3 (2.99\% AO) but performance drops substantially on Candor-LR (26.63\% AO). Adding LRS2 to training improves results slightly on Candor-LR (22.72\% AO). Training on Candor-LR alone yields much better Candor-LR performance, with AO WER falling to 14.81\% and AV WER to 10.50\%. Lastly, the combined LRS2+LRS3+Candor-LR training gives the best overall Candor-LR result, reaching 9.83\% AV WER. This setup also maintains strong performance on LRS2 (2.54\% AV) and LRS3 (1.42\% AV).

The visual benefit ($\Delta$) on Candor-LR is smallest when trained on scripted data alone (+1.77 LRS3-only, +1.40 LRS2+LRS3) but rises sharply once Candor-LR is included in training (+4.31 Candor-LR, +6.54 combined). This shows that visual cues become substantially more important once the model is exposed to realistic conversational data, rather than simply scaling scripted data volume. In contrast, LRS3 visual gains remain stable across all training setups, showing that adding Candor-LR to training does not hurt out-of-domain performance on LRS3.

\subsection{Noise Robustness for Candor-LR}
\label{sec:snr}

\begin{figure}[t]
\vspace{-0.5em}
\centering
\begin{subfigure}[b]{0.48\textwidth}
    \caption{LRS3 testset (1 hour)}
    \includegraphics[width=\linewidth]{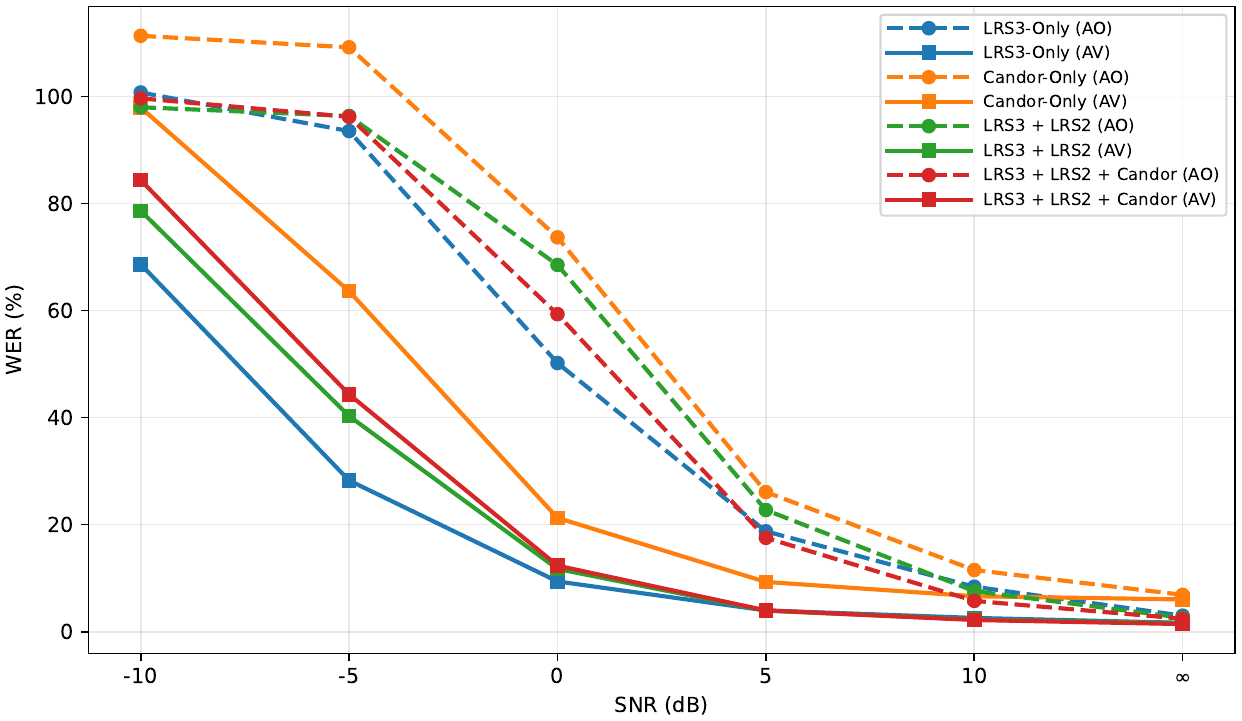}
    \label{fig:lrs3_matched}
\vspace{-1.5em}
\end{subfigure}

\begin{subfigure}[b]{0.48\textwidth}
    \caption{Candor-LR testset (60 hours)}
    \includegraphics[width=\linewidth]{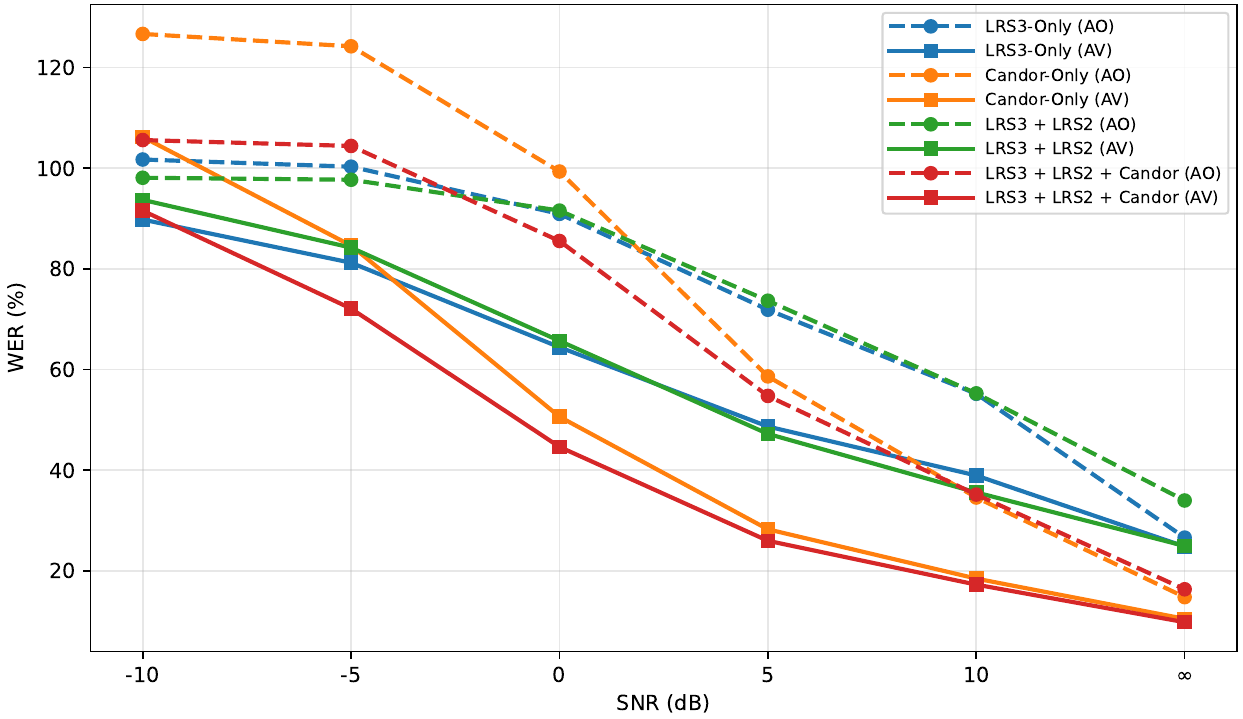}
    \label{fig:candor_matched}
\end{subfigure}
\vspace{-1.5em}
\caption{WER across SNR levels for LRS3 and Candor-LR test sets when \textbf{trained without noise augmentation}, where $\infty$ dB denotes clean speech and different colors indicate training data combinations.}
\label{fig:snr_clean}
\vspace{-1.5em}
\end{figure}

\begin{figure}[t]
\vspace{-0.5em}
\centering
\begin{subfigure}[b]{0.48\textwidth}
    \caption{LRS3 testset (1 hour)}
    \includegraphics[width=\linewidth]{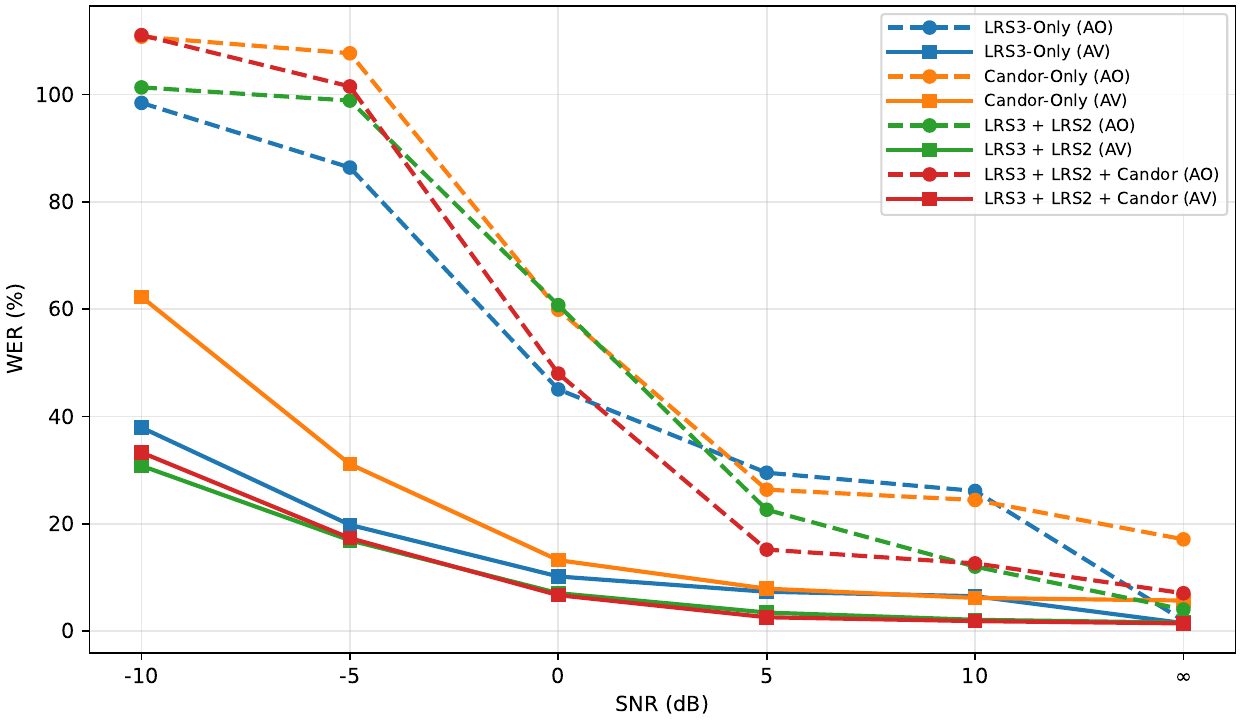}
    \label{fig:candor_on_lrs3}
\vspace{-1.5em}
\end{subfigure}

\begin{subfigure}[b]{0.48\textwidth}
    \caption{Candor-LR testset (60 hours)}
    \includegraphics[width=\linewidth]{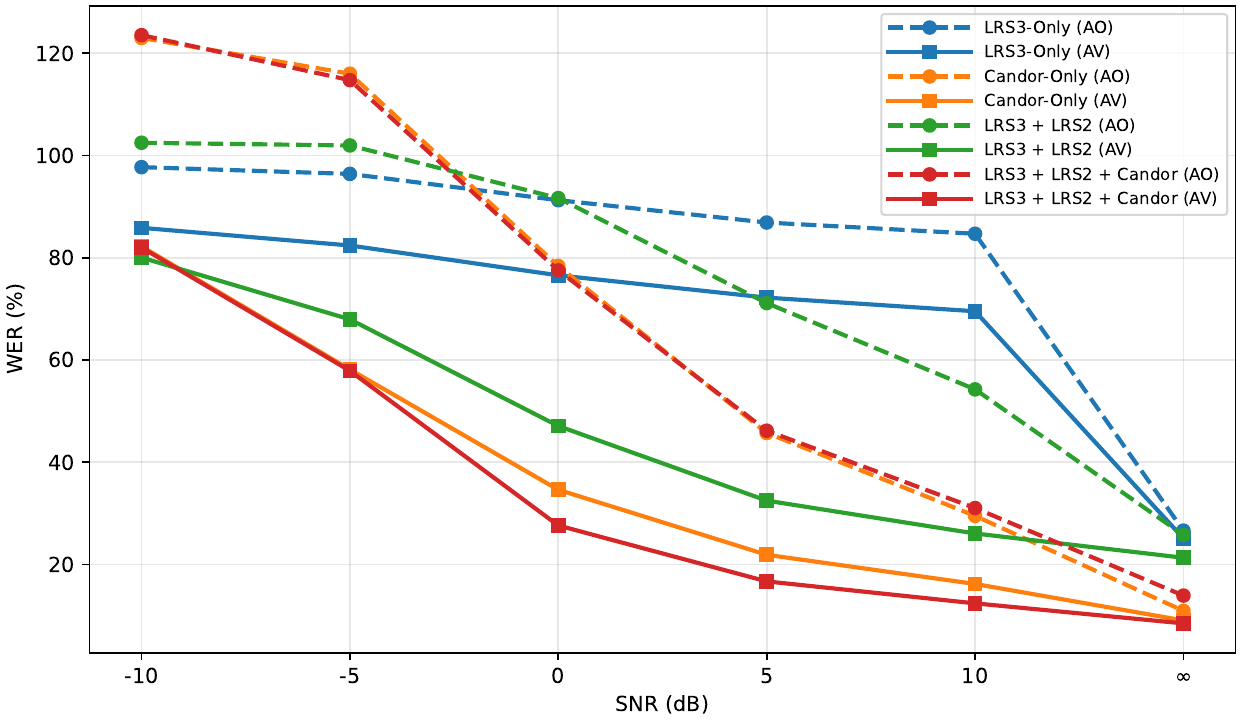}
    \label{fig:lrs3_on_candor}
\end{subfigure}
\vspace{-1.5em}
\caption{WER across SNR levels for LRS3 and Candor-LR test sets when \textbf{trained with noise augmentation}, where $\infty$ dB denotes clean speech and different colors indicate training data combinations.}
\label{fig:snr_noisy}
\vspace{-1.5em}
\end{figure}

To better understand the characteristics of the Candor-LR dataset under challenging acoustic conditions, we evaluate its performance trends under noise and compare them against the standard LRS3 benchmark. Using AV-HuBERT as our baseline model, we augment both datasets with synthetic babble noise. The noise generation follows the same procedure used by the AV-HuBERT authors for LRS3 \cite{avhubert, shi22_interspeech}, where babble noise is synthesized from multiple overlapping speech sources from training data and added at different signal-to-noise ratio (SNR) levels ($-10$dB to $+10$dB). We experiment with two setups: applying this noise augmentation only at inference (with clean training), and incorporating it during both training and inference. Figure~\ref{fig:snr_clean} and Figure~\ref{fig:snr_noisy} plot the WER performance across SNR levels for the inference-only noise and joint train-and-test noise configurations, respectively, across both the LRS3 and Candor-LR test sets. While performance predictably drops on both datasets as noise increases, the widening gap between them highlights the unique difficulty of Candor-LR.

\subsubsection{Evaluation with Inference-Only Noise}
\label{subsec:inference_only_noise}

As shown in Figure \ref{fig:snr_clean}, WER decreases for both datasets as SNR increases, but Candor-LR shows a higher sensitivity to noise. In completely clean conditions (no added noise), the AO LRS3 model achieves a 2.99\% WER on the LRS3 test set, while the Candor-LR model records a 14.81\% WER on Candor-LR. At 10 dB SNR, LRS3 degrades slightly to 8.39\% WER, whereas Candor-LR increases to 34.60\% WER. At -10 dB SNR, LRS3 reaches 100.73\% WER, while Candor-LR reaches 126.66\% WER, indicating high insertion errors. This indicates that Candor-LR is inherently more vulnerable to noise.

Introducing the visual modality provides consistent improvements, but examining the gap between AO and AV performance reveals that visual cues become increasingly important as noise increases. For models trained exclusively on in-domain data (i.e., trained and tested on the same corpus), incorporating the visual modality at 10 dB SNR reduces the Candor-LR WER by 16.11\% (34.60\% AO vs 18.49\% AV). The gap widens at 0 dB SNR, where speech and noise are equally loud. At this level, the AO model on Candor-LR reaches a 99.33\% WER, while the AV model maintains a 50.63\% WER, highlighting a 48.7\% WER difference bridged by visual data. At -10 dB SNR, visual modalities become necessary for Candor-LR to mitigate severe degradation. Expanding the training dataset to the combined LRS2+LRS3+Candor-LR corpus further validates this observation. This combined dataset yields a 17.29\% AO WER on the Candor-LR test set at 10 dB. Yet, under the severe -10 dB condition, the model exhibits a 91.56\% AO WER, showing that diverse, large-scale data alone cannot replace structured exposure to acoustic noise during training.

Cross-dataset generalization is highly asymmetric. Training exclusively on Candor-LR and testing on LRS3 yields a 6.62\% AV WER at 10 dB (and 6.02 in clean conditions). Conversely, training on LRS3 and testing on Candor-LR yields a 38.96\% AV WER at 10 dB. While the larger volume of training data in Candor-LR naturally contributes to this improved generalization, it also demonstrates that Candor-LR's combined scale and naturalistic distribution effectively encompass the more controlled LRS3 domain. 

\subsubsection{Evaluation with Joint Train-and-Test Noise}
\label{subsec:joint_noise}

Building on this sensitivity to noise observed in Figure \ref{fig:snr_clean}, we next examine whether exposing the model to noise during training can mitigate these effects. Following the noise augmentation approach used in AV-HuBERT, our second configuration trains with noise, randomly mixing babble noise into the training data with a 25\% probability. This noise-trained setup substantially narrows the AO-AV performance gap observed earlier, as shown in Figure \ref{fig:snr_noisy}. Most notably, augmentation increases the model's reliance on visual information precisely when acoustic conditions deteriorate. For the LRS2+LRS3 model evaluated on the LRS3 test set at -10 dB, the AV WER drops from 78.62\% in the clean-trained setup to 30.79\% in the noise-trained setup, a relative improvement that far exceeds anything achieved by scaling data alone. A similar, though smaller, gain appears on Candor-LR, where AV WER improves from 93.70\% to 80.03\% under the same conditions. This result suggests that noise exposure during training does more than make the model robust to noise; it teaches the model to rely on noise-invariant visual features, compensating for audio degradation that clean-trained models cannot handle.

Interestingly, the data reveals a penalty for applying noise augmentation to narrow, single-domain datasets. When the LRS3-only model is evaluated on Candor-LR at 10 dB, the clean-trained version achieves a 38.96\% WER, but the noise-trained version degrades sharply to 69.53\%. This suggests that adding synthetic noise to a limited dataset causes the model to overfit to that specific acoustic distribution, actively hurting its ability to generalize to out-of-domain, naturalistic speech.

Expanding the training data to the full LRS2+LRS3+Candor-LR corpus removes this penalty and reveals a more nuanced relationship between data diversity and noise tolerance. At 10 dB SNR, combining diverse training data with noise augmentation drops the LRS3-only AV WER from 6.48\% to a near-perfect 1.84\%. For the Candor-LR test set, however, moving from the Candor-LR model to the combined model brings only a modest gain at 10 dB, from 16.19\% to 12.40\% AV WER. The smaller gain on Candor-LR indicates that scripted data diversity helps LRS3 more than it helps Candor-LR at moderate noise levels.

As noise increases to the -10 dB extreme, though, diverse training data combined with noise augmentation becomes critical for avoiding total model failure. The Candor-LR model trained without noise augmentation reaches a 106.20\% AV WER at -10 dB, while the fully optimized model (combined data and noise training) reduces this to 82.00\% AV WER. In-domain data is therefore necessary for handling unscripted conversational speech, but diverse out-of-domain data paired with noise augmentation becomes essential once noise interference gets severe. Overall, Candor-LR proves far tougher than LRS3 as a benchmark for evaluating model robustness in real-world conditions.

To provide a definitive baseline for future research, we summarize the best performance on the Candor-LR test set across all acoustic conditions in Table \ref{tab:benchmark_snr}. 

\begin{table}[t]
\centering
\caption{Benchmark WER (\%) comparison across SNR levels on the Candor-LR test set (trained on LRS2+LRS3+Candor-LR). $\infty$ represents clean audio.}
\label{tab:benchmark_snr}
\footnotesize
\setlength{\tabcolsep}{4pt}
\renewcommand{\arraystretch}{1.2}

\begin{tabularx}{\columnwidth}{|l| *{6}{>{\centering\arraybackslash}X|}}
\hline
\rowcolor{headerblue}
& \multicolumn{6}{c|}{\textbf{SNR (dB)}} \\ \cline{2-7}
\rowcolor{headerblue}
\textbf{Modality} & \textbf{-10} & \textbf{-5} & \textbf{0} & \textbf{5} & \textbf{10} & \textbf{$\infty$} \\
\hline
VO & \multicolumn{6}{c|}{69.50} \\ 
\hline
AO & 123.54 & 114.74 & 77.53 & 46.17 & 31.04 & 13.92 \\ 
\hline
AV & 82.00 & 57.88 & 27.63 & 16.69 & 12.40 & 8.48 \\
\hline
\end{tabularx}
\vspace{-1em}
\end{table}

\section{Conclusion}
In this work, we provide an open-source pipeline to derive AVSR-ready data from the CANDOR dataset. Using this pipeline, we present Candor-LR, a large-scale AVSR benchmark for conversational settings built from natural dyadic videoconference data. It contains 713.5 hours of training data, 10.1 hours of validation data, and 60.1 hours of test data, capturing spontaneous turn-taking speech and realistic acoustic conditions that are largely absent from scripted datasets. We benchmark pretrained AVSR models and evaluate cross-domain generalization under varying training configurations and noise conditions. Our results show that current AVSR models struggle with unscripted, conversational speech, with visual modality and in-domain exposure playing a key role in narrowing this gap. Candor-LR offers a challenging real-world benchmark that moves AVSR beyond scripted datasets toward conversational speech understanding.

\section*{AI-Generated Content Disclosure}
Gemini was used during the preparation of this work for minor grammatical edits and to enhance the clarity of the writing.

\bibliographystyle{IEEEtran}
\bibliography{my_bib}

\end{document}